\documentstyle[12pt]{article}
\begin{document}
\vspace*{6cm}

\begin{center}
{\large Comments on ``Finsler Geometry and Relativistic Field Theory"}
\end{center}

\vspace*{.5cm}

\begin{center}
{\bf{Shervgi S. Shahverdiyev$^*$}}
\end{center}
\begin{center}
{\small   Institute of Physics, Azerbaijan National Academy of Sciences, Baku, Azerbaijan}
 \end{center}
\vspace{.5cm}

\begin{center}
{\bf{Abstract}}
\end{center}

\begin{quote}
\noindent
{\small We show that results obtained in paper {\it{Foundations of Physics 33, No. 7, 1107 (2003)}}, are not correct.}
\end{quote}

\vspace{1cm}

\noindent

\noindent

\vspace{5cm}

\hspace{.1cm}${}^*$e-mail:shervgis@yahoo.com

http://www.geocities.com/shervgis

\newpage

In a recent paper \cite{b}, R. G. Beil claims to have derived
electromagnetism from the so called Finsler geometry. However,
this result contradicts one of the results of \cite{sh}, where it
is shown that
 electromagnetism can not be geometrized in the framework of Riemannian geometry with any metrics.
 Because Finsler geometry is Riemannian geometry with Finsler metric \cite{ssc},
 electromagnetism cannot be geometrized in the its framework too.

In the present paper
 we demonstrate that results obtained in \cite{b} are not correct.
In section 4 of \cite{b},  R. G. Beil, after choosing matrix (52) gets connection of Riemannian geometry (59)
$$
\Gamma_{\lambda\mu\nu}=\frac{1}{2}k\left[B_\lambda(\frac{\partial B_\mu}
{\partial x^\nu}+\frac{\partial B_\nu}{\partial x^\mu})\right]+
\frac{1}{2}k\left[B_\mu(\frac{\partial B_\lambda}{\partial x^\nu}-
\frac{\partial B_\nu}{\partial x^\lambda})+B_\nu(\frac{\partial B_\lambda}
{\partial x^\mu}-\frac{\partial B_\mu}{\partial x^\lambda})\right].(59)
$$
Then he imposes condition (60)-(61)
on field $B$
$$
B_\nu\upsilon^\nu=\frac{e}{mck}, \quad \frac{\partial
B_\nu}{\partial x^\mu}\upsilon^\nu= \frac{\partial}{\partial
x^\mu}(B_\nu\upsilon^\nu)=0. \quad \quad \quad \quad \quad
\quad\quad \quad \quad \quad \quad\quad\quad\quad(60)
$$
Using this condition and (59) he represents equation of motion
$$
\frac{d\upsilon_\lambda}{d \tau}+\Gamma_{\lambda\mu\nu}\upsilon^\mu\upsilon^\nu=0
$$
in the form (62)
$$
\frac{d\upsilon_\lambda}{d \tau}+\frac{e}{mc}(\frac{\partial B_\lambda}
{\partial x^\mu}-\frac{\partial B_\mu}{\partial x^\lambda})\upsilon^\mu=0.
 \quad\quad\quad \quad\quad \quad\quad \quad \quad \quad \quad \quad
   \quad\quad\quad \quad \quad \quad\quad(62)
$$
However, after replacing (59) to equation of motion and imposing (60) we obtain
$$
\frac{d\upsilon_\lambda}{d \tau}+\frac{e}{mc} \frac{\partial
B_\lambda}{\partial x^\mu}\upsilon^\mu=0 \quad\quad\quad
\quad\quad \quad\quad \quad \quad \quad \quad \quad\quad \quad
\quad \quad \quad \quad \quad\quad \quad\quad\quad (1)
$$
instead of (62). It is easy to see this  by calculating
$$\Gamma_{\lambda\mu\nu}\upsilon^\mu\upsilon^\nu=\frac{1}{2}k\left[B_\lambda(\frac{\partial B_\mu}{\partial x^\nu}+\frac{\partial B_\nu}{\partial x^\mu})+B_\mu(\frac{\partial B_\lambda}{\partial x^\nu}-\frac{\partial B_\nu}{\partial x^\lambda})+B_\nu(\frac{\partial B_\lambda}{\partial x^\mu}-\frac{\partial B_\mu}{\partial x^\lambda})\right]\upsilon^\mu\upsilon^\nu=
$$
$$
\frac{1}{2}k\left[B_\lambda\upsilon^\nu\frac{\partial B_\mu}{\partial x^\nu}
\upsilon^\mu+B_\lambda\upsilon^\mu\frac{\partial B_\nu}{\partial x^\mu}\upsilon^\nu+
B_\mu\upsilon^\mu\frac{\partial B_\lambda}{\partial x^\nu}\upsilon^\nu-B_\mu
\upsilon^\mu\frac{\partial B_\nu}{\partial x^\lambda}\upsilon^\nu+
B_\nu\upsilon^\nu\frac{\partial B_\lambda}{\partial x^\mu}\upsilon^\mu-
B_\nu\upsilon^\nu\frac{\partial B_\mu}{\partial x^\lambda}\upsilon^\mu\right]=
$$
$$
k\left[B_\lambda\upsilon^\nu\frac{\partial B_\mu}{\partial x^\nu}\upsilon^\mu+
B_\mu\upsilon^\mu\frac{\partial B_\lambda}
{\partial x^\nu}\upsilon^\nu-B_\mu\upsilon^\mu\frac{\partial B_\nu}{\partial x^\lambda}
\upsilon^\nu\right]
$$
and imposing (60)
$$
\Gamma_{\lambda\mu\nu}\upsilon^\mu\upsilon^\nu\left|_{B_\nu\upsilon^\nu=\frac{e}
{mck},\frac{\partial B_\nu}{\partial x^\mu}\upsilon^\nu=0}\right.=
$$
$$
k\left[B_\lambda\upsilon^\nu\frac{\partial B_\mu}{\partial
x^\nu}\upsilon^\mu+ B_\mu\upsilon^\mu\frac{\partial
B_\lambda}{\partial
x^\nu}\upsilon^\nu-B_\mu\upsilon^\mu\frac{\partial B_\nu}{\partial
x^\lambda}\upsilon^\nu\right]_{\left|_{B_\nu\upsilon^\nu=\frac{e}{mck},\frac{\partial
B_\nu}{\partial x^\mu}\upsilon^\nu=0}\right.}=
\frac{e}{mc}\frac{\partial B_\lambda}{\partial x^\nu}\upsilon^\nu
$$
which give (1) instead of (62).

Next, R. G. Beil compares (62) with equation of motion
$$
\frac{d\upsilon_\lambda}{d \tau}+\frac{e}{mc}F_{\mu\lambda}\upsilon^\mu=0
\quad\quad \quad\quad \quad \quad \quad \quad\quad \quad  \quad\quad \quad \quad
 \quad  \quad\quad\quad \quad \quad \quad \quad\quad(2)
$$
and claims that
$$
F_{\mu\lambda}=\frac{\partial B_\lambda}{\partial
x^\mu}-\frac{\partial B_\mu} {\partial x^\lambda}. \quad
\quad\quad \quad \quad\quad \quad \quad \quad \quad \quad \quad
\quad \quad \quad\quad  \quad\quad\quad \quad \quad \quad
\quad(63)
$$
This claim is not correct because by comparing (62) with $(2)$
we obtain
$$
F_{\mu\lambda}\upsilon^\mu=(\frac{\partial B_\lambda}{\partial
x^\mu}-\frac{\partial B_\mu}{\partial x^\lambda})\upsilon^\mu.
\quad\quad \quad\quad \quad \quad \quad \quad \quad \quad
\quad\quad \quad\quad \quad \quad \quad \quad \quad \quad\quad(3)
$$
Basic solutions to this equation cannot be (63) because $\upsilon^\mu$ are not independent due to condition (60).
In order to solve $(3)$ we have to count condition (60) which gives
$ F_{\mu\lambda}=\frac{\partial B_\lambda}{\partial x^\mu}$. (63) is a linear combination of these basic solutions. Note that linear combination of basic solutions in the form $\frac{\partial B_\lambda}{\partial x^\mu}+\frac{\partial B_\mu}{\partial x^\lambda}$ is also a solution to (3) because of (60).

After (63), R. G. Beil claims that $F_{\mu\lambda}$ can be
identified with electromagnetic field. However, any tensor in the
form of (63) can not be identified with electromagnetic field,
because in order $F_{\mu\lambda}$ to be identified with
electromagnetic field  it must satisfy Maxwell equations
$\partial_\mu F_{\mu\lambda}=0$.

From the physical point of view condition (60) is not acceptable for electromagnetic field and charged particles, because it expresses charge of a particle as a function of its velocity and electromagnetic potential. It is very well known that charge of a particle depends neither on its velocity nor on the potential of external electromagnetic field.  Also, (60) gives electromagnetic potential as a function of inverse velocities, which is not acceptable as was pointed out in \cite{sh}.

We also would like to note that the so called Finsler geometry is not a geometry different from Riemannian geometry. It is actually Riemannian geometry with the so called Finsler metric (see for example \cite{ssc}).
R. G. Beil (as many authors) uses notions of Finsler geometry and Finsler metric interchangeably.

I would like to stress that as it is proven in \cite{sh},
Riemannian geometry with any metric is not suitable for
geometrization of electromagnetism because there is no equivalence
principle for electromagnetic interaction.

Finally, in sec. 5, R. G. Beil states that metric with explicit
appearance of $e/m$ is introduced by G. Randers. However, this is
not the case. In his work \cite{r}, G. Randers considered
Riemannian geometry with metric without any coefficients. Metric
and new geometries with explicit appearance of $e/m$ have been
introduced in \cite{sh}.

\end{document}